# MACHO Messages from the Big Bang

## G. Chapline and J. Barbieri

The present day mass spectrum for dark matter compact objects is calculated based on the assumption that a uniform population of PBHs was created at a definite red-shift, and that the mass spectrum evolved as a result of gravitational radiation. The predicted present day spectrum extends over many decades of mass and allows one to connect the abundance of MACHOs in the halo of our galaxy with the abundance of galactic seeds. Present day astrophysical constraints on the abundance of dark matter PBHs appear to be consistent with our predicted mass spectrum if it is assumed that the seeds for the present day dark matter MACHOs were created at a time ~ $10^{-4}$ second after the big bang. Remarkably the total cosmological energy density at this time obtained by extrapolating the sum of the present day dark matter and CMB energies backward in time is very close to the mass-energy density of an Einstein-de Sitter universe at the same time. This suggests that the radiation precursor to the CMB was created at about the same time as the seeds for the present day dark matter.

*Introduction* Some time ago it was suggested [1,2] that the dark matter component of the matter in today's universe might consist entirely of primordial black holes (PBHs). It was also pointed out [3] that in a flat universe whose mass-energy density is dominated by primordial black holes it would perhaps be natural for all matter to eventually consist of horizon mass BHs. Of course these proposals begged the question as to why dark matter should consist of black holes rather than some other exotic form of non-baryonic matter such as WIMPs or axions. On the other hand the persistent failure to find any evidence for WIMPs or cosmic axions has perhaps tipped the balance in favor of PBHs [4,5]. In this letter we focus on the question of what the present day spectrum of dark matter compact objects might have to say about the form of matter near to the onset of the big bang. We show that if an initially uniform population of PBHs evolves as a result of gravitational radiation during binary collisions, then the present day spectrum for primordial compact objects will smoothly interpolate between the MACHO objects which could form the halo of our galaxy [6] and the massive seeds for the massive compact objects at the centers of galaxies [7,8,9]. This puts constraints on the redshift where the initial PBHs were formed, which naturally leads to the question as to what form matter took prior to this redshift. An intriguing possibility is that the PBH precursors to the present day MACHOs and the radiation precursor to CMB were both created at the same red-shift by the release of entropy from massive compact objects created at the onset of the big bang.

*Predicted dark matter mass spectrum.* Our basic assumption is that the present day population of dark matter MACHOs evolved from an initial population of compact objects with nearly the same mass $M_{DM}$ created at a specific redshift $z_r$. We will also assume that during the radiation dominated era $z < z_r$ the evolution of cosmological parameters for the universe proceeds as in the standard cosmological model [10]. In addition we assume that the time corresponding to the redshift $z_r$ occurs before the time of cosmological nucleosynthesis, since the creation of precursor to present day dark matter might be expected to affect the standard model nucleosynthesis predictions which are in excellent agreement with observations. With these assumptions we have used a Boltzmann-like equation to numerically calculate the dark matter MACHO mass spectrum that would evolve during the radiation era as result of binary collisions between compact objects where gravitational radiation is important. In particular we assume that the time-dependent probability density p(M,$t$) that describes how the fraction of primordial MACHOs with a mass between M and M+dM changes with time is mainly due to gravitational radiation induced coalescence of MACHO pairs, and this increase is described by a Boltzmann-like equation

$$\frac{dp(M)}{dt} = v(t)\frac{\rho_{DM}}{\bar{M}}\int_{M_{DM}}^{\infty} \pi b_{cap}^2 \, p(M')p(M-M')dM' \qquad (1)$$

where $\rho_{DM}$ is the dark matter mass density at time t, $v(t)$ is the relative velocity of the compact objects at time t and $b_{cap}(M',M-M')$ is the critical impact parameter that would allow two passing compact objects with masses M' and M-M' to be captured into a stable orbit via gravitational radiation [11]. We believe that radiative capture is the most important mechanism for shaping the mass spectrum since the critical impact parameter for radiative capture is typically much larger than the Schwarzchild radius. In Eq.1 time can run from the time corresponding to the redshift $z_r$ (viz. ~ $10^{-4}$ sec) to the present time. However we have found that the present day MACHO mass spectrum is largely fixed in the first few minutes after the big bang; i.e. by the time cosmic helium was produced. What to use for the average relative velocity $v(t)$ in Eq.(1) is of course problematic. As a rough approximation we have used the "virial" $v = [(1+z)/10^{12}]^{1/2}10^5$km/sec ; i.e. we assume $v^2(t)$ varies roughly as the as the inverse of the distance between dark matter particles, and reaches typical halo velocities by the time of matter dominance $z < 10^4$. Some sample solutions to Eq 1, assuming that the initial mass distribution $p(M,z=z_r)$ is concentrated at $M = M_{DM}$, are shown in Fig.1.

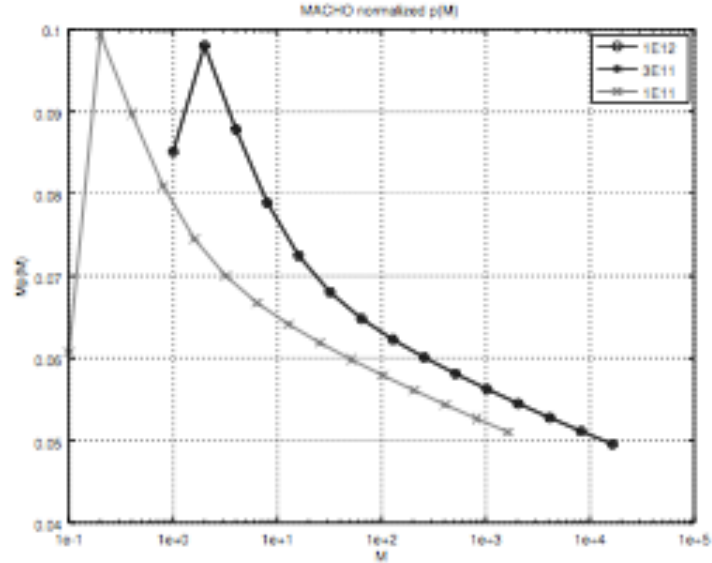

Fig. 1 Dark matter mass spectra for $M_{DM}$ = 0.1 and $1 M_\odot$ and $z_r$ = $10^{12}$ and $3 \times 10^{11}$ corresponding to initial times $t_r \sim 10^{-5} - 10^{-4}$ sec.

Although we could use Eq. (1) to calculate the mass spectrum for any combination of initial mass $M_{DM}$ and the red shift $z_r$, we have focused on the spectrum of compact objects resulting from collisions between primordial compact objects with initial masses, $M_{DM}$, in the range 0.01 – 1 $M_\odot$ and values of $z_r$ > !0$^{10}$ . Figure 1 shows 2 examples of such calculations, which assumed $M_{DM}$ = 0.1 and 1 $M_\odot$. It can be seen that the spectrum is fairly flat, and that no logarithmic interval in mass accounts for more than ~10% of the dark matter density. This makes our prediction based on these initial masses consistent with previous direct searches for dark matter compact objects in the halo of our galaxy using gravitational micro-lensing [12,13]. As well as widely accepted astrophysical constraints on what typical MACHO masses might be [14,15]. Our predicted spectrum also extends up to $10^4$ solar masses with sufficient strength to possibly explain the origin of galaxies [7,8,9].

It is the flatness of our predicted mass spectrum that may allow our predicted dark matter mass spectrum to evade previous claims (see [14,15] that massive compact objects are excluded as candidates for dark matter. Indeed it was recently noted [16] noted that a lognormal mass spectrum would effectively negate the claim that a dark matter density of primordial compact is inconsistent with searches for micro-lensing of supernovae. The mass spectra shown in Fig 1 also appear to evade the astrophysical constraints discussed in Ref 14 (cf. Fig. ) if one interprets to constraints shown in Ref. 14 for a monochromatic spectrum as constraints on the integral of our probability density over some logM interval, say $\Delta logM = 1$. Astrophysical constraints on an extended mass spectra are also discussed in Ref. [15]. On the positive side there already may be direct evidence for the existence of primordial compact objects. Partly as a result of the unexpected observation of gravitational radiation from the coalescence of 2 compact objects with estimated masses near to 30 solar masses, there is considerable interest at the present time in the question whether the compact objects observed by LIGO could be primordial in origin [17]. Even if the LIGO objects are primordial though, the mass $M_{DM}$ is still undetermined because the LIGO objects may have formed by coalescence of smaller compact objects. The particular choices for $M_{DM}$ shown in FIG. 1 were chosen keeping in mind that the relative abundance of dark matter objects with masses $\sim 10^4 M_\odot$ should explain the observed abundance of galaxies. In particular, we assume that our predicted mass spectrum should conform with current estimates [7,8,9] that the galactic seeds should represent $\sim 10^{-4}$-$10^{-5}$ of the dark matter density.

*What was the form of matter prior to $10^{-4}$ sec?* Since the conventional view is that BHs are indestructible what physical meaning can be ascribed to our assumption that present day dark matter evolved from a population of PBHs created at a time ~ $10^{-4}$ sec? The answer to this question may lie with the answer to the puzzles as to what physical process could have led to the radiation precursor to the CMB and how does one explain the specific entropy of the CMB?. In principle inflation theories can produce radiation via reheating [10] or dissipation [18], and hybrid inflation models are capable of producing massive PBHs [19,20]. However to our knowledge inflation theories have to date not provided any explanation for the specific entropy of the CMB as measured by the number of CMB photons per gram of dark matter. This number is nearly the same at the red-shift $z_r$ as today, and so the mystery is really to not just explain the sudden appearance of PBHs but also explain why these PBHs were accompanied by a specific number of photons and lepton pairs.

Our proposal for resolving this enigma is based on the observation that if $M_{DM}$ lies in the range ~ 0.1-1$M_\odot$, then the estimated radiation temperature at $z = z_r$ is very close to the temperature where, due to quantum effects, the surface of the compact object would no longer no longer transparent to the radiation as it would be in classical general relativity. Indeed it has been predicted [21] that due to quantum effects when the photon energy reaches a critical value ≈ 300 MeV$(M_\odot /M)^{1/2}$ it will interact strongly with the surface modes of the compact object. This would allow compact objects to interact strongly with radiation, and as a consequence of the very large heat capacity of such objects a large density of such objects would transform ambient radiation into entropy stored in the surface modes of the compact objects. As a corollary one can make an independent estimate for the critical red-shift $z_r$

$$(1+z_r)h\nu_{CMB} \approx 0.3 \text{ GeV}(M_\odot/M_{DM})^{1/2} \ . \tag{2}$$

This estimate is consistent with the estimates based on MACHO mass spectra that are consistent with astrophysical constraints. Thus we are led to the hypothesis that prior to the red-shift $z_r$ the matter in the universe consisted almost exclusively of black holes, rather than a mixture of PBHs and radiation.

Curiously, if we extrapolate the present day energy density of the CMB (.026eV/cm$^3$) to a red-shift ~ $10^{12}$, then the resulting energy density $\rho_r$ is close to the energy density of an Einstein – de Sitter universe at the same epoch. In other words the total matter mass-energy density of our universe at the redshift $z_r$ is apparently very nearly the same as in a flat Robertson–Walker universe containing only cold matter. This means that prior to $z_r$ cosmological matter could have consisted entirely of BHs. Serendipitously today's observed density of dark matter is also completely consistent with the hypothesis that the PBH precursors to todays dark matter MACHOs arose from the decay of a massive compact objects.

In the scenario for the production of massive PBHs described in ref. [3] it was assumed that has at some very early epoch the universe is dominated by matter with a Frautschi-Hagedorn equation of state (which is the appropriate equation of state for a gas of black holes). Very massive compact objects are then continuously formed as a result of the large density fluctuations due to Poison fluctuations in a universe dominated by initially close packed BH. Indeed the BH dominated universe described in ref. [3] has the property that for any sphere with radius $R_0$ where there is a positive increase $\delta\rho$ in mass density – no matter how small the increase in density – the sphere will not expand indefinitely, but instead will reach a maximum radius $R_{max} \approx R_0(\rho/\delta\rho)$, where the density inside the sphere is approximately 3 x the ambient mass density. Subsequently the sphere will collapse to form a BH. All the compact objects formed during this Einstein-de Sitter-like epoch will have a mass close to $M_H$ because any compact masses with a smaller mass will get "swept up" into the collapse of the largest objects formed by the collapse of the mass within the horizon $\approx 0.5(ct)$. Combining Eq. (2) with the observed energy densities for the CMB (.026 eV/cm³) and dark matter density ($\approx$ 1keV/cm³), yields an estimate for mass of the horizon scale BHs mass at $z = z_r$ :

$$M_H(z_r) = 20 \left[\frac{1+z_r}{10^{12}}\right]^4 M_\odot \qquad (3)$$

Although other explanations for the origin of the present day MACHOs and the CMB may eventually emerge, the simplicity of the idea that both present day spectrum of MACHOs and the CMB are the result of the release at $z = z_r$ of internal energy by BHs with the mass (3) is attractive. This scenario also provides a prediction for the specific entropy of the CMB [22]. This predicted entropy was not inevitable but is a consequence of the large difference between the horizon mass (3) and the residual mass $M_{DM}$ estimated from the current constraints on the MACHO mass spectrum. This scenario relies on the very large heat capacity of compact of compact objects, and provides a natural explanation for the somewhat mysterious circumstance that at the present time matter consists of a component with large entropy (the CMB) and a dark matter component with seemingly very low entropy (discounting the unobservable Hawking entropy). We also note that as a result of the sweeping up of all matter into massive BHs prior to $z = z_r$ no primordial WIMPS, SUSY particles, monopoles, or other exotic elementary particles will survive into the observable era z < $z_r$.

During the epoch z > $z_r$ we hypothesize that cosmological matter will consist mostly of massive BHs. Since the cosmological pressure due to BHs is negligible compared to the mass density, the mass-energy density for z > $z_r$ will vary in the familiar way when mass is conserved ρ = $ρ_r$[(1+z)/(1+ $z_r$ )]³ . Combining this with the relation between time and red-shift for a Robertson-Walker universe with zero pressure matter $t/t_r \approx$ [(1+ $z_r$)/(1+z)]³/² provides an estimate for the horizon scale masses prior to z = $z_r$ :

$$M_H = \left[\frac{1+z_r}{1+z}\right]^{\frac{3}{2}} M_H(z_r) \qquad (4)$$

We see that the horizon mass decreases as z increases, and so the mass-energy density would diverge as $z \to \infty$ if only horizon scale BHs were present. However if the horizon scale BHs were the result of coalescence of smaller BHs, then the mass-energy density will have a finite maximum $ρ_*$ which is attained when these smaller BHs become close packed. In a sense all the large scale features of our universe are determined by value of this $ρ_*$. It is perhaps worth mentioning that this scenario is similar to Zeldovich's cosmological model [23], where he assumed that the universe began with close packed nucleons. On the other hand, our mechanism for the production of the CMB is rather different from Zeldovich's.

*What is the initial energy density?* As a result of renewed efforts to use gravitational micro-lensing to detect compact objects with masses > 1 solar mass it may not be too far off before we know for sure what the present day MACHO spectrum looks like. Combining this knowledge with solutions to Eq. (1) one can then determine the total mass-energy density $\rho_r$ at red-shift $z_r$. Unfortunately the initial energy density $\rho_*$, corresponding to the time when the initial PBHs for the $z > z_r$ era become close packed, is undetermined. However, the distinct possibility that the initial PBHs may originated at a finite red-shift $z_* \gg z_r$ suggests that the emergence of our expanding universe is associated with a phase transition of the vacuum state at some energy density $\gg (GeV)^4$. It is tempting to speculate that this initial energy density may be related to the breaking of supersymmetry. Finally we note that our scenario does not preclude an inflationary episode proceeding $z_*$. However from our point of view that our observed universe began with a universe filled with BHs, it is perhaps more natural to note [24] that the "flatness" problem can also be solved by assuming that the era $z_* > z > z_r$ was proceeded by an era where the universe was flat and contained matter, but the vacuum energy was negative.

## Acknowledgments

The authors are very grateful for discussions with Carlos Frenk, Paul Frampton, Michael Schneider, Will Dawson, and Nathan Golovich.

This work was performed under the auspices of the U.S. Department of Energy by Lawrence Livermore National Laboratory under Contract DE-AC52-07NA27344 and was supported by the LLNL-LDRD Program under Project No. 17-ERD-120.